\documentclass[a4paper,12pt]{amsart}

\usepackage[cp1251]{inputenc}
\usepackage[english]{babel}
\usepackage{epsfig}
\usepackage{amssymb}
\usepackage[all]{xy}

\sloppypar

\begin{document}

\title{MODEL OF THE EXPANSION \\
OF H {\small II} REGION RCW 82}

\author{K.V. Krasnobaev}
\address{Lomonosov Moscow State University\\
Faculty of Mechanics and Mathematics\\
119991 Moscow, Russia\\
Russia Space Research Institute of RAS\\
117997 Moscow, Russia}
\email{kvk-kras@list.ru}
\author{R.R. Tagirova}
\address{Russia Space Research Institute of RAS\\
117997 Moscow, Russia}
\email{rtaghirova@gmail.com}
\author{G.Yu. Kotova}
\address{Lomonosov Moscow State University\\
Faculty of Mechanics and Mathematics\\
119991 Moscow, Russia}
\email{gviana2005@gmail.com}

\begin{abstract}
This paper aims to resolve the problem of formation of young objects observed in the RCW 82 H {\small II} region. In the framework of a classical trigger model the estimated time of fragmentation is larger than the estimated age of the H {\small II}  region. Thus the young objects could not have formed during the dynamical evolution of the H {\small II}  region. We propose a new model that helps resolve this problem. This model suggests that the H {\small II}  region RCW 82 is embedded in a cloud of limited size that is denser than the surrounding interstellar medium. According to this model, when the ionization--shock front leaves the cloud it causes the formation of an accelerating dense gas shell. In the accelerated shell, the effects of the Rayleigh--Taylor (R-T) instability dominate and the characteristic time of the growth of perturbations with the observed magnitude of about 3 pc is 0.14 Myr, which is less than the estimated age of the H {\small II}  region. The total time $t_{\sum}$, which is the sum of the expansion time of the H {\small II}  region to the edge of the cloud, the time of the R-T instability growth, and the free fall time, is estimated as $0.44<t_{\sum } < 0.78$ Myr. We conclude that the young objects in the H {\small II}  region RCW 82 could be formed as a result of the R-T instability with subsequent fragmentation into large-scale condensations.

{\it Key words}: {H {\small II} regions -- hydrodynamics -- instabilities --  ISM: individual objects (RCW 82) -- stars: formation }

\end{abstract}

\maketitle

\section{INTRODUCTION}
Observations by \cite{Pom09}  show that the Galactic H {\small II} region RCW 82 is in a state of active star formation. They discussed the possible mechanism of formation of the observed young objects in the framework of a trigger model  \cite{EL77}. According to this model, due to the expansion of the H {\small II}  region, a complex of discontinuities called the ionization--shock front appears. It includes an ionization front and a prior shock wave \cite{Spit78, Ost06}. The distance between the front and the shock increases in time. The mass of the gas compressed by the shock wave increases with time. When the integral mass of the layer between the ionization front and the shock wave exceeds the Jeans mass, fragmentation of the layer into smaller scale condensations occurs \cite{Whith94}. In the case where the ionization--shock front propagates in a homogeneous gas, the estimated time of fragmentation will be approximately 1.6 Myr according to Pomares et al. (2009). At the same time, the estimated age of the H ii region RCW 82 is approximately 0.4 Myr, thus the young objects could not have formed during the dynamical evolution of the H  {\small II} region.
 
However, there is another possible way for large-scale condensations in the ionization--shock front layer to form. For this we need to assume that the medium is not homogeneous and the H {\small II} region is embedded in an interstellar cloud of limited size, which is denser than the surrounding interstellar medium. When the ionization--shock front leaves the cloud an accelerated dense gas shell is formed  \cite{Kot09}. The acceleration of the shell opens the possibility of Rayleigh--Taylor (R-T) instability developing and as a consequence makes the formation of large-scale condensations possible.

The influence of R-T instability on the appearance of nonuniform structures in the outer parts of H {\small II}  regions and on the formation of gas clumps has been considered by many authors. \cite{Cap73} and \cite{Cap06} showed that the origin of the drop-shaped condensations, which were observed in the nebula NGC 7293, could be due to R-T instability development. They considered the case where the scale of the condensations was smaller than the thickness of the shell. \cite{Mal87} presented evidence that the origin of clumps around $ \lambda $ Orionis could be caused by R-T instability. According to \cite{Reip97}, the globules observed in the H {\small II} IC 2944 could	appear at	the late stage of R-T	instability development. \cite{Bar77} and \cite{Kras04}  investigated the stability of ionization fronts in accelerating gas. \cite{Giul79} found that the non-stationary ionization--shock front motion is unstable to long-wave perturbations. \cite{Sch80}  showed that R-T instability can also cause fragmentation of the dense shell swept out by the stellar wind in the H {\small II}  region. \cite{Miz05, Miz06} modeled the plane flow caused by the action of ionizing radiation on an isolated layer of neutral gas. These authors used models that do not take into account the cooling caused by inclusions of impurity elements, the heating by dissociation, and the spectral radiation transfer. The model of the two-dimensional motion of H {\small II} region formation in a spherical cloud, which took into account the contribution of the impurity elements in the heat balance and the dependence of the absorption coefficient on frequency, was considered by \cite{Kot09, Kot10}. They found that in the thin layer between the shock and ionization fronts finger-shaped condensations, containing a large part of the initial perturbation masses, can appear. \cite{Kras13} constructed a linear stability theory of the self-gravitating layer, which is accelerated by the pressure difference on both of its sides, and gave the classification of the instability regimes, which depend on the relation between the gravitational force and the pressure force. \cite{Wal13}  considered the evolution of a H {\small II} region formed in a molecular cloud with a given fractal dimension. It was found that either a few massive clusters or many objects with smaller masses: columns (pillars) and globules, depending on the fractal dimension, can possibly form.

In the H {\small II} regions isolated condensations surrounded by ionized gas are frequently observed. Their origin may be associated with the compression and ionization of pre-existing inhomogeneities (e.g., clumps, globules, columns, etc.). In particular, the equilibrium configuration of globules has been studied by several authors \cite{Kahn69, Dys73}. The dynamic influence of converging shocks on globules in one-dimensional motion was considered by \cite{Dib64}. \cite{Dys75} proposed the model of axisymmetric flow around the globule.  \cite{Lefl97} used the model of clump compression by the ionization--shock front to interpret the observations of the globule in the H {\small II} region IC 1848.  \cite{Bod79} and \cite{ Mell98} calculated the interaction between an ionization--shock front and a dense cloud. They found that the clump has a complicated form, which differs from the drop shape. However, in the case of H  {\small II}  RCW 82, the model of compression of the pre-existing clumps cannot sufficiently explain the typical location of the condensations, which are only on the periphery of the H  {\small II}  regions, and the absence of condensations inside them. Furthermore, there are no data indicating interaction between the gas of the H {\small II} regions and the gas heated by the radiation of the central star flowing away from the surface of the clump.

Since the structure of clumps depends on the conditions of their formation, it is possible to compare the observational data of the clump parameters with the theoretical models of the evolution of the dense layer on the boundary of the H  {\small II} region. In our paper we propose a model of R-T instability in a shell that is generated by the ionization--shock front and accelerated as it goes out of a cloud with limited size and into the interstellar medium. \cite{Kot09, Kot10}, using the framework of a complete system of equations for radiation gas dynamics, developed the following features of the evolution of H {\small II} regions in a spherical cloud. They found that if the cloud radius does not substantially exceed an initial Str\"{o}mgren radius, the typical stages of expansion of the H  {\small II} region are as follows. In the first the ionization front rapidly propagates; hydrodynamic motion is insignificant. The second stage is characterized by the appearance of an ionization--shock front and a dense layer between these discontinuities. In the third stage the shock wave is going out of the cloud border and an accelerating shell is formed. In the latter stage a rarefaction wave appears in the shell, which causes the smothering of the density distribution and the reduction of the contrast between the density of the cloud and the environment with lower density.

It was found that the duration of the acceleration stage is sufficient for R-T instability development. Despite the fact that the density of the expanding H {\small II} region decreases, the layer acceleration is maintained by the effect of the ``rocket'' force of the gas flowing from the ionization front. This conclusion agrees with the simple estimates of the acceleration value  \cite{Cap73, Cap06, Kras08}. Deceleration of this layer is important in the case when either the mass of the shocked matter is large (i.e., the cloud radius is large) or when the gas density outside the cloud is high enough.
 
The plan of this paper is as follows. In Section 2 a mathematical model of the accelerated motion of the ionization--shock front layer is described. Section 3 presents the results of the evolution of the perturbations of the layer. The morphology of the growing perturbations is studied, and the integral masses of the condensations are defined. The influence of self-gravitation and geometry of the motion on these condensations is estimated. The application of our model to the observations of the H {\small II}  region RCW 82 \cite{Pom09} is presented in Section 4. We show that according to our model the mass accumulation of the condensations leads to a decrease of the Jeans length and accordingly to reduction of the time of its gravitational compression. As a result, the estimated total time $t_{\sum }$, which consists of the time of the expansion of the H {\small II}  region RCW 82 to the border of the cloud, the time of R-T instability development, and the time of gravitational compression, belong to the interval $0.44<t_{\sum } < 0.78$ Myr.

\section{MODEL}

To formulate the model we make two simplifying assumptions. The first simplifying assumption made in our model is the adiabatic behavior of the flow. This approach is valid only in the case of perturbations that increase quickly enough.

The second simplifying assumption is the replacement of the ionization front by tangential discontinuity. This is possible because at the hydrodynamic stage of expansion of the H {\small II}  region, the ionization front is a D-type front, which is characterized by small pressure changes at the sides of the front \cite{Spit78}.

We consider unsteady motions of the compressible self-gravitating gas. The equations expressing conservation of mass, momentum, and entropy and the Poisson equation for the gravitational potential are
\begin{equation}
\begin{split}
&\frac{\partial \rho}{\partial t}+
 \textrm{div}  \rho {\bf v} {=0}, \\
&\rho \, \frac{d \bf{v}}{d t}=- \textrm{grad} \, p - \rho \, \textrm{grad} \, U, \\
&\frac{d}{d t} \biggl (\frac{p}{\rho^{\gamma}} \biggr )=0, \\
& \Delta U=4 \pi G \rho,
\end{split}
\label{bcs}
\end{equation}
where $\rho$, ${\bf v}$, and $p$ are the mass density, the velocity vector, and the pressure, respectively.  Here $\gamma$, $G$,  and  $U$ are the adiabatic index, the gravitational constant, and the gravitational potential, respectively.

To describe the two-dimensional motion we introduce  the coordinates ($x_{1 }\, $, $x_{2}$), which are  Cartesian ($x, y$), cylindrical ($r,z$), and spherical ($r,\vartheta$)  (where $\vartheta$ is the polar angle).  The velocity vector has two components $u$ and $\upsilon$ in the $x_{1}$ and $x_{2}$ directions, respectively.

\begin{figure}
\begin{center}
\includegraphics[scale=0.7]{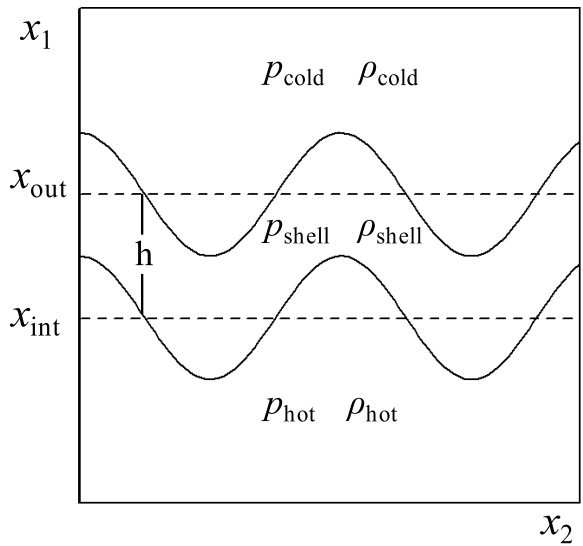}
\end{center}
\caption{Variables that define the initial condition. The ionized gas in the H {\small II} region is characterized by the pressure $p_{\rm hot}$  and density $\rho_{\rm hot}$. At $t=0$, when the ionization-shock front reaches the cloud boundary, we assume that the gas parameters between the fronts  are $p_{\rm shell}$  and  $\rho_{\rm shell}$. The perturbed and unperturbed shell is drawn by solid and by dotted line, respectively. The interstellar medium have $p_{\rm cold}$  and  $\rho_{\rm cold}$}
\label{fig1}
\end{figure}

We start our modeling at the time-moment  $t=0$, which is when the shock wave reaches the boundary of the cloud. Though the distribution of the gas velocity in the H {\small II}  region has a complex shape, the detailed distribution of the gas weakly influences the shell impulse in time (and consequently on the acceleration; \cite{Kot09}). Therefore we assume that the gas is at rest (i.e. ${\bf v}=0$ at $t=0$), and the density and the pressure are determined as follows:
\begin{equation}
\begin{split}
a<x_1<x_{\rm int}: \,  \,   \rho=\rho_{\rm hot}, \, p=p_{\rm hot} , \\
x_{\rm int}<x_1<x_{\rm out}: \,  \,  \rho=\rho_{\rm shell}, \, p=p_{\rm shell} , \\
x_{\rm out}<x_1<b: \,  \,  \rho=\rho_{\rm cold}, \, p=p_{\rm cold} .
\end{split}
 \label{bcs}
\end{equation}
Figure \ref{fig1} shows the initial condition.
The surface $x_1=x_{\rm int}$ corresponds to the ionization front, which is replaced in our model by the tangential discontinuity and therefore $p_{\rm hot}=p_{\rm shell}$. The region $a<x_1<x_{\rm int}$ is occupied by a hot gas. $x_1=x_{\rm int}$ is the inner boundary of the high density layer. The surface $x_1=x_{\rm out}$ is the outer boundary of the high density layer with initial thickness $h={\rm const}$,  $x_1=x_{\rm out}$ is the end of cloud (i.e. for  $x_1 > x_{\rm out}$ the gas is rarefied) and the shock reaches this boundary at $t=0$. Therefore, we have an arbitrary discontinuity there at time $t=0$, where  $p_{\rm shell} > p_{\rm cold}$. The gas in the outer region ($x>x_{\rm out}$) is assumed to have very low temperature and density. Both $\rho_{\rm hot}$ and $\rho_{\rm cold}$ are significantly smaller than $\rho_{\rm shell}$, which means that   the gas in the shell has a higher density than the gas outside the shell. The distribution of the temperature between the shock and the ionization front has a complex shape due to the multiplicity and variety of cooling mechanisms \cite{Bod79, Miz06, Whal08, Kot09, Iw11}.  However, the details of the temperature distribution do not change the integral characteristics of the resulting motion we are interested in. Therefore we assume that the gas in the shell has the temperature of the same order of magnitude  as  the gas outside the shell (i.e $p_{\rm shell}/\rho_{\rm shell} \backsim p_{\rm cold}/\rho_{\rm cold}$).
We consider the  periodic (by the coordinate $x_2$) perturbed parameters of the shell. We set the type of perturbations, which are close to those considered by \cite{Zon03} in a study of the effects  of the mass accumulation in the thin deformed shells. Therefore, at $t=0$, for $x_1$: $x_{\rm int}<x_1<x_{\rm out}$ we assume fluctuations of velocities:
\begin{equation}
0<x_2<\lambda/2: \, \, u \,^{\prime} =A \, \cos(2 \pi x_2/\lambda), \, \upsilon \,^{\prime}
=A \, \sin(2 \pi x_2/\lambda)\, ,
 \label{bcs}
\end{equation}
where $A$ and $\lambda$  are the amplitude and  wavelength. In spherical coordinates ($r, \vartheta$)  the expression for $x_2$ in Equation (3) is changed by $x_2 =\vartheta x_{\rm int}$.

The boundary conditions are as follows: for the values $\rho$, $p$, and ${\bf v}$ in the system of Equations (1) we assume that the boundaries $x_1=a$ and $x_1=b$ are  impermeable, i.e., $u=0$ \cite{Dud99}, and periodic by the coordinate $x_2$.  For the potential $U$ (or gravitating field) we set boundary conditions everywhere along the boundary surface of the computational domain. Those values are obtained analytically from the integral expression \cite{Tikh90} for the external potential (or gravitating field).  In the case of Cartesian coordinates at boundaries $x_1$ we set inhomogeneous Neumann boundary conditions for ${\partial U}/{\partial x_1}$; at penetrable boundaries  $x_2$ we use symmetrical  conditions, i.e., ${\partial U}/{\partial x_2}=0$. We consider the self-gravity for a plane problem, because in this case it is easy to obtain analytically the periodic boundary conditions for the gravitational field. For the other coordinates, the self-gravity  will be  considered in future papers.

We solve the equations of gas dynamics (Equation(1)) numerically  with the so-called total variation diminishing Lax-Friedrichs scheme \cite{Hof00, Kul01}. In order to obtain the potential in the inner computational domain we solve the Poisson equation by means of  the discrete Fast Fourier Transform method.

\section{TWO-DIMENSIONAL PERTURBATIONS AND MASS ACCUMULATION}
In the works of \cite{Kras08} and \cite{Kot10}, the sufficient conditions for the appearance of the regular finger-shaped condensations with mass accumulation were established. It is shown that the condensations appear when the density of the shell significantly exceeds the density of the hot gas (more exactly, when the ratio $\rho_{\rm shell}/\rho_{\rm hot} \gtrsim 10$) in the presence of long wave perturbations $h \ll \lambda$. In our calculations the initial conditions (Equations (2) and (3)) were chosen to satisfy these criteria. We also chose the parameters of the shell in such a way that the ratio of the mass of the hot and the cold gas roughly corresponds to that which is observed in a typical H {\small II} region.

To estimate the role of self-gravity, it is convenient to introduce the parameter $\beta$, defined as the ratio of the characteristic value of	the gravitational acceleration in the layer	$2 \pi G \rho_{\rm shell} h$ to the acceleration	$W=(p_{\rm hot}-p_{\rm cold})/(\rho_{\rm shell} h)$ caused	by the pressure difference on two sides of the dense shell: \[\beta=2 \pi G \rho_{\rm shell} h/W.\]

The one-dimensional calculations  \cite{Kras13}  show that for $\beta \rightarrow 0$ and $\rho_{\rm hot}$, $\rho_{\rm cold} \rightarrow 0$ self- gravity effects are not essential, and the motion of the gas is characterized by the formation of several discontinuities with a subsequent acceleration of the layer. In our problem the density  $\rho_{\rm shell}$ considerably exceeds  $\rho_{\rm hot}$ and $\rho_{\rm cold}$; nevertheless the mass of the hot ionized gas contained in the Str\"{o}mgren zone is large and is comparable with the mass of the layer. This fact, as well as the presence of cold gas before the layer, makes gravitational effects important for the gas motion.

In the problem considered it is convenient to normalize the pressure, the density, the velocity, and the temperature ($p_{\rm hot}$, $\rho_{\rm hot}$, $u_{\rm hot}=\sqrt{p_{\rm hot}/\rho_{\rm hot}}$ and $T_{\rm hot}$) by their values in the region of the  hot gas. The spatial scale is normalized by some arbitrary value $L$. We assume that the typical values of the H {\small II} region are $\rho_{\rm hot}=1.67 \times 10^{-22}$ g cm$^{-3}$, $T_{\rm hot}=10^4 K$,  $p_{\rm hot}=2.76 \times 10^{-10}$ erg cm$^{-3}$, $u_{\rm hot}=12.9 \, \text{km} \, \text{s}^{-1}$. The scale  $L=1$ pc is chosen to be the average size of the condensations on the borders of the H  {\small II} region, and the characteristic time is equal to  $t_0=L\sqrt{\gamma-1}/u_{\rm hot}=0.06$ Myr.

\begin{figure}
\begin{center}
\includegraphics[scale=0.6]{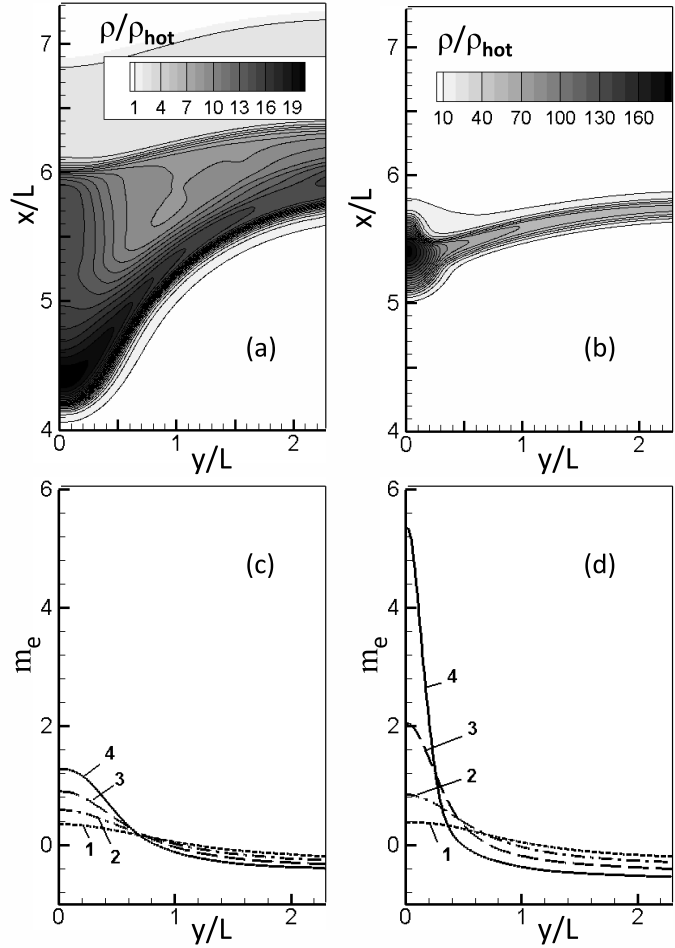}
\end{center}
\caption{Morphology of the condensations for the plane layer without self- gravity ((a), (c)) and taking it into account self-gravity effects ((b), (d)) with ${\beta=3.1}$. Panels (a) and (b) show density isolines at the time $t/t_0=5$. Panels (c) and (d) show the function $m_{\rm e}(y)$ at $t/t_0=2,3,4,5$ (curves 1--4). The perturbation parameters are $A\sqrt{\gamma-1}/L=0.1$ and $\lambda/L=5$}
\label{fig2}
\end{figure}

General properties of the two-dimensional motion of the self-gravitational  layer are shown in Figure \ref{fig2} with the initial conditions $\rho_{\rm shell}/\rho_{\rm hot}=25$, $p_{\rm shell}/p_{\rm hot}=1$, $\rho_{\rm cold}/\rho_{\rm hot}=1$, $p_{\rm cold}/p_{\rm hot}=0.04$, $h/L=0.5$, $\gamma=5/3$.
In this simulation the boundaries of the computational domain are  $a=0$, $b/L=10$ and the initial position of the shell is $x_{\rm int}/L=4.75$.

Numerical modeling of two-dimensional motions shows that at the nonlinear stage of evolution of the perturbation, the effects of R-T instability dominate for $\beta < 1$ (Figure \ref{fig2}(a)).

If the gravitational force has the same order of magnitude as or larger than the pressure difference on both sides of a layer, i.e. $\beta > 1$, then the structure of the condensations changes and the differences between the sizes of the inhomogeneities in the longitudinal and transverse directions are significantly reduced, as shown in Figure \ref{fig2}(b) for $\beta=3.1$. Self-gravity leads to a significant increase of the maximum of the density in the self-gravitating layer as compared to the case of R-T instability in the model without self-gravitation.

Let us consider the quantitative characteristics of mass accumulation as the integral value of $m_{\rm e}(x_2,t)$:
$$m_{\rm e}(x_2,t)=\frac{M_{\rm S}-M_{\rm S}^{\ast}}{M_{\rm e}^{\ast}}, \, M_{\rm S}(x_2,t)=
\int_{a}^{b} \rho (x_1,x_2,t)dx_1 \, ,$$
$$M_{\rm S}^{\ast}(0)=\int_{a}^{b} \rho (x_1,0,0)dx_1 \, ,
M_{\rm e}^{\ast}(0)=\int_{x_{\rm int}}^{x_{\rm out}} \rho (x_1,0,0)dx_1 .$$

The function $m_{\rm e}$ is shown in Figures \ref{fig2}(c) and (d). The maximum of the integral masses in the self-gravitating layer at $t/t_0=5$ exceeds the value of $m_{\rm e}$ for $\beta \rightarrow 0$ several times. The volume occupied by the dense gas decreases accordingly.
\begin{figure}
\begin{center}
\includegraphics[scale=0.6]{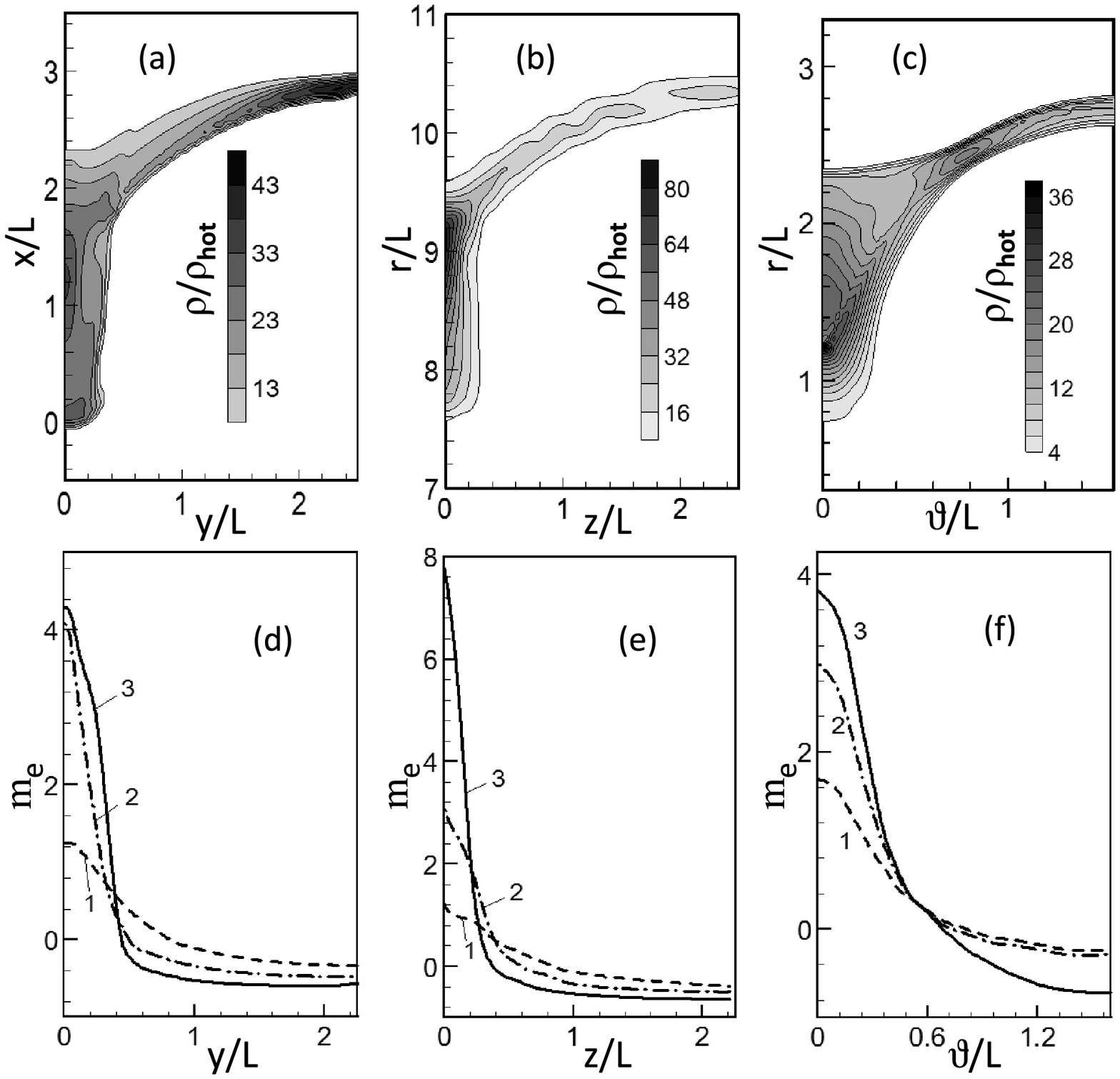}
\end{center}
\caption{Morphology of the condensations for a plane layer (a,d), and cylindrical (b,e) and spherical (c,f) shells. The density isolines are shown at time $t/t_0=8$  (a, b) and $t/t_0=6$ (c). Plots (d, e, f) correspond to $m_{\rm e}$ for $t/t_0=4,6,8$ (curves 1--4)}
\label{fig3}
\end{figure}

The influence on the considered problem of the geometry of the motion (e.g., plane, spherical or cylindrical) is illustrated in Figure \ref{fig3} for the case with no self-gravitation. In these simulations the initial conditions were close to those discussed in Figure \ref{fig2}; the other values were changed to  $\rho_{\rm shell}/\rho_{\rm hot}= 50$,  $p_{\rm cold}/p_{\rm hot}=0.02$, $h/L=0.25$. The amplitude of perturbations is equal to $A/L=0.1$. The wavelength for both plane and cylindrical geometries was chosen as previously $\lambda/L=5$, and for the spherical coordinate is presented for $\lambda/L=\pi$. In Figure \ref{fig3} the time $t$ is normalized by $t_0=L/u_0$ and the value of the initial position of the ionization front $x_{\rm int}/L$ is equal to $0$, $8.0$ and $1.5$  for the plane layer, and the cylindrical and the spherical shell, respectively.

The results presented in Figure \ref{fig3} show that mass accumulation takes place for all types of motion (plane, cylindrical, or spherical) for the case when the density of the gas in the shell significantly exceeds the density of the hot gas. The area of the mass concentration is narrow enough and considerably smaller than the wavelength of the perturbations. Therefore we have shown here that the structure of the condensations and the effect of mass accumulation in the shell slightly depend on the geometry of the problem.

\section{INTERPRETATION AND CONCLUSIONS}

Numerical simulations allow us to estimate the characteristic time of layer destruction, which depends on the scale of perturbations and the parameter $\beta$.  We will use the results of our calculations to identify the influence of the acceleration on the appearance of large-scale condensations (clumps) in the external parts of the H {\small II} region RCW 82. According to the observations of \cite{Pom09}, the present radius  of the H {\small II} region $R_{\rm i}$ is equal to 3 pc. The scale of  condensations near the boundary separating the neutral and the ionized gas is also on the order of 3 pc. The mass  of the nebula $M_{\rm n}$ is approximately $2520 \,M_{\odot}$, and the mass of the shell $M_{\rm e}$ is approximately $2260 \,M_{\odot}$. \cite{Pom09} suggested that the H {\small II} region RCW 82 could have been formed under the influence of flux ionizing radiation of $\Phi=9 \times 10^{48} \text{s}^{-1}$ in a homogeneous medium with a concentration of neutral particles $n_{0}=10^{3} \text{cm}^{-3}$ (additionally the authors of this work give other values of the parameters of the gas and the  flux $\Phi$ which differ slightly from previous values). Then the age of  the H {\small II} region $t_{\rm n}$ is approximately $0.4$ Myr. This is less than the time of the fragmentation, which is equal to $1.6$ Myr. Therefore, the young objects (which are the possible candidates for star formation regions) observed in the H {\small II} region RCW 82  are not formed according to the model by \cite{Pom09}, since due to this model the objects have to appear much later than $t_{\rm n}$.  

However, if we assume that the H {\small II} region was not formed in a homogeneous medium but in a cloud with mass $M_{\rm n}$ and with number density $n_{0}=10^{3} \text{cm}^{-3}$, then the accelerating neutral shell could be formed during the time $t_{\rm c}$, which is considerably less than $t_{\rm n}$. Indeed, let $R_{\rm c}$ be the initial cloud radius, which is considered to be spherical and to consist of atomic hydrogen for simplicity. Then we have $4\pi m_{\rm H} n_{0}R_{\rm c}^{3}/3=M_{n}$ and, therefore, $R_{\rm c}=2.92$ pc (where $m_{\textrm{H}}$ is the mass of a hydrogen atom). The corresponding Str\"{o}mgren radius $R_{\rm S}$ is equal to
\[
R_{\rm S}=\left ( \frac{3 \Phi}{4 \pi n_{0}^{2} \alpha_{\textrm{H}}} \right
)^{1/3}= 0.71 \, \, \text {pc} \, ,
\]
where $\alpha_{\textrm{H}}=2 \times 10^{-13} \text{cm}^{3} \textrm{s}^{-1}$ is the coefficient of photo recombination at all levels of a hydrogen atom, except the ground level.
The radius $R_{\rm S}$ of the static H {\small II} region is reached during  the time $(n_{0} \alpha_{\textrm{H}})^{-1} \ll t_{n}$ \cite{Spit78}. Therefore the time $t_{\rm c}$, when
 the ionization front passes the distance $R_{\rm c}$ (i.e. reaches the boundary of the cloud, forming accelerated moving shell) can be defined by  the approximate formula \cite{Bar77, Spit78}
\begin{equation}
\frac{R_{\rm c}- R_{\rm S}}{R_{\rm S}}=\left ( 1+\frac{7}{4} \frac{c
t_{\rm c}}{R_{\rm S}} \right )^{4/7}.
 \label{f:5}
 \end{equation}
Here $c$ is the  isothermal  sound speed in a fully ionized hydrogen gas  at the temperature  $T_{\rm e}=10^{4}K$. Note that Equation (4) agrees well with the results of the modeling of an ionization--shock front leaving the cloud \cite{Kot09}. Using Equation  (4) we find that $t_{\rm c} \approx 0.23$ Myr. Assuming that the  luminosity of the star is invariable we see that the number density of the ionized gas $n_{\rm e}$ and the radius $R_{\rm i}$ of the H {\small II} region are related by $n_{\rm e} R_{\rm i}^{3/2}=\textrm{const}$. Thus
\[
n_{\rm e}(t_{\rm c})=n_{0}\biggl(\frac{R_{\rm S}}{R_{\rm i}(t_{\rm c})} \biggr )^{3/2}
\approx 120 \, \text{cm}^{-3},
\]
which agrees with the value given in the work of \cite{Pom09}.
Assuming that $n_{\rm e}=120 \text{cm}^{-3}$, we estimate the acceleration of the shell $W$ and the parameter $\beta$
\[
\begin{split}
&W=\frac{8 \pi R_{\rm i}^{2}n_{\rm e} kT_{\rm e}}{2260  M_{\odot}}=8
\times
10^{-8}  \, \, \textrm{cm s}^{-2}\,  , \\
&\beta=\frac{2 \pi G \rho h}{W}=\frac{G M_{\rm e}}{2 W R_{\rm i}^{2}}=0.02
\end{split}
\]
(where $\rho h$ is the surface density of the shell and $k$ is the Boltzmann constant).

During the acceleration of the shell the effects of R-T instability dominate and the characteristic time $\tau$ of the growth of perturbations with scale  $R_{\rm i}=2 \pi /\kappa$ is $\tau=(W \kappa)^{-1/2} \approx 0.14 $ Myr.

Thus, at the time of $t_{\rm c}+\tau=0.37$ Myr  the condensations could be formed in the layer whose integrated density exceeds the integrated density of the unperturbed layer several times. The time of development of Jeans instability in these clumps has the same order as the free fall time $t_{\rm ff} =  \sqrt{3 \pi/ (32G \rho_{\rm av})}$   \cite{Spit78},  where $\rho_{\rm av}$  is the average value of the layer density. If the layer mass $M_{\rm e}$ is known, then the $\rho_{\rm av}$  is determined by the detailed structure of the layer -- i.e., its  thickness and the spatial distribution of the density. Numerous calculations  \cite{Ten79, Miz05, Kot09} show that the thickness and the density distribution depend considerably on the abundances of the impurity elements in the neutral gas, on its degree of ionization, and on the spectra of the star (or the stars group) that excite the H {\small II} region. However, we can estimate the minimum and maximum values  of $\rho_{\rm av}$  and  therefore  the value of $t_{\rm ff}$.  

The maximum value of  $\rho_{\rm av}$  corresponds to isothermic flow in the  layer  and it can be roughly estimated from the condition of equality between the pressure in the layer  and the pressure in the H {\small II} region. Assuming that in the HII region we have $n_{\rm e}=120 \text{cm}^{-3}$ and  $T_{\rm e}=10^4 K$ we obtain that $2 n_{\rm e} k T_{\rm e}=\rho_{\rm av} a_{\rm av}^2$, where $a_{\rm av}=0.2 \, \text{km} \, \text{s}^{-1}$ is the chaotic average particle velocity in the layer \cite{Pom09}. Then  the minimum of  $t_{\rm ff} $ is equal to $ 0.07$ Myr. 

To estimate the minimum value of  $\rho_{\rm av}$  we assume that after the shock wave goes out of the border of the cloud, the value of the density of the compressed neutral gas is not less than four times of the unperturbed density $\rho_{0}$, since the flow is adiabatic with the index $\gamma = 5/3$. This value $4 \rho_{0}$ is the minimum value, since after the shock front  gas radiative cooling takes place  that result in a density increase. If we take into account that the clumps density increased approximately four times due to the development of  R-T instability (referring to Figure  \ref{fig3}, one can see that $m_{\rm e}$ may reach the value of $m_{\rm e} \sim 4$), it is reasonable to take $\rho_{\rm av} = 16 \rho_{0}$. Then we have  the maximum  of  $t_{\rm ff} \approx 0.407$ Myr. 

Consequently, the time of formation of young objects $t_{\sum}=t_{\rm c}+\tau+t_{\rm ff}$ is estimated as $0.44<t_{\sum } < 0.78$ Myr.  

Thus the surface density of the shell increases and the time of gravitational compression decreases. Therefore, the young objects in the Galactic H {\small II} region RCW 82 could be formed as the result of the development of R-T instability with subsequent fragmentation into large-scale condensations. For a detailed analysis of the process of fragmentation one needs to make a more complete assessment of the effects of radiation heating and cooling and examine an essentially nonlinear stage of growth of perturbations in the self-gravitating gas.
 
We would like to thank V. Izmodenov for useful discussions. This work has been supported by the RFBR grants 11-01-00409 and 14-01-00747.


\begin{thebibliography}{}
\bibitem[Baranov \&  Krasnobaev (1977)] {Bar77} Baranov, V.B. \&  Krasnobaev,  K.V., 1977, Hydrodynamic Theory of Space Plasma (Moscow: Nauka) [in Russian]
\bibitem[Bodenheimer et al. (1979)]{Bod79} Bodenheimer, P., Tenorio-Tagle, G. \& Yorke, H.W., 1979, ApJ, 233, 85
\bibitem[Capriotti (1973)]{Cap73}  Capriotti, E. R., 1973, ApJ, 179, 495
\bibitem[Capriotti \& Kendall (2006)]{Cap06}  Capriotti, E. R.   \& Kendall, A. D., 2006, ApJ, 642, 923 
\bibitem[Dibai \& Kaplan (1964)]{Dib64} Dibai, E. A. \&  Kaplan, S. A., Astronomicheskii Zhurnal, 1964, 41, 652
\bibitem[Dudorov et al. (1999)] {Dud99} Dudorov, A. E.,  Zhilkin, A. G. \&  Kuznetsov, O. A., 1999, Mathematical Models and Computer Simulations, 11:1, 101
\bibitem[Dyson (1973)]{Dys73} Dyson, J. E., 1973, A\&A, 27, 459 
\bibitem[Dyson (1975)]{Dys75} Dyson, J. E., 1975, Ap\&SS, 35, 299.
\bibitem[Elmegreen \&  Lada (1977)]{EL77} Elmegreen, B.G. \&  Lada, C.J., 1977, ApJ, {214}, 725
\bibitem[Giuliani (1979)]{Giul79} Giuliani, J.L., 1979, ApJ, 233, 280
\bibitem[Hoffmann \& Chiang (2000)]{Hof00} Hoffmann, K.A. \& Chiang, S.T., 2000, Computational fluid dynamics  (Wichita: Engineering Education Systems 4th ed.)
\bibitem[Iwasaki et al. (2011)]{Iw11} Iwasaki, K.,  Inutsuka, S. \&   Tsuribe T., 2011, ApJ, {733}, article ID. 17
\bibitem[Kahn (1969)]{Kahn69} Kahn, F. D., 1969, Physica, 41, 172
\bibitem[Kotova  \& Krasnobaev (2009)]{Kot09} Kotova, G.Yu. \&  Krasnobaev, K.V., 2009, Astronomy Letters, 35, 167
\bibitem[Kotova  \& Krasnobaev (2010)]{Kot10} Kotova, G.Yu. \&  Krasnobaev, K.V. 2010, Astronomy Letters, 36, 479
\bibitem[Krasnobaev  (2004)]{Kras04} Krasnobaev, K. V., 2004,  Astronomy Letters, 30, 451
\bibitem[Krasnobaev \& Tagirova (2008)]{Kras08} Krasnobaev, K. V. \& Tagirova, R. R., 2008,  Fluid Dynamics, {43}, 814
\bibitem[Krasnobaev \& Tagirova (2013)] {Kras13} Krasnobaev, K. V. \& Tagirova, R. R., 2013,  Astronomy Letters, 39, 578
\bibitem[Kulikovsky et al. (2001)]{Kul01} Kulikovsky, A.G., Pogorelov, N.V., Semenov, A.Y. 2001 Mathematical Problems of the Numerical Solution of Hyperbolic Systems of Equations (Moscow: Phizmathlit) [in Russian]
\bibitem[Lefloch et al. (1997)]{Lefl97} Lefloch, B., Lazareff, B. \&  Castets, A., 1997, ApJ, 324, 249
\bibitem[Mellema et al. (1998)]{Mell98} Mellema, G., Raga, A. C., Canto, J.,  et al.,  1998,  A\&A,  331, 335 
\bibitem[Malone et al.  (1987)]{Mal87} Malone, D., McBreen,  B. \& Fazio, G. G., 1987, Irish Astronomical Journal, 18, 91
\bibitem[Mizuta et al. (2005)]{Miz05}  Mizuta, A., Kane, J. O.,  Pound, M. W., Remington, B.A.,  Ryutov, D.D. \&  Takabe, H., 2005, ApJ, 621, 803
\bibitem[Mizuta et al. (2006)]{Miz06} Mizuta, A.,  Kane, J.O.,  Pound, M.W.,  Remington, B.A.,   Ryutov, D.D. \&  Takabe, H., 2006, ApJ, {647}, 1151
\bibitem[Osterbrock (2006)]{Ost06} Osterbrock, D. E. \& Ferland, G. J. 2006, Astrophysics of Gaseous Nebulae and Active Galactic Nuclei (California: Univers. Sci. Books, 2-nd ed.)
\bibitem[Pomares  et al. (2009)]{Pom09} Pomares, M., Zavagno, A.,  Deharveng, L.,  et al., 2009, A\&A, {494}, 987
\bibitem[Reipurth et al. (1997)]{Reip97} Reipurth, B., Corporon, P.,  Olberg, M.  \&  Tenorio-Tagle, G., 1997, A\&A, 327, 1185
\bibitem[Schneps et al. (1980)]{Sch80} Schneps, M.H., Ho, P.T.P.  \&  Barrett, A.H, 1980, ApJ, 240, 84
\bibitem[Spitzer (1978)]{Spit78} Spitzer, L., 1978, Physical Processes in the Interstellar Medium (New York: Wiley)
\bibitem[Tenorio-Tagle  et al. (1979)]{Ten79} Tenorio-Tagle, G., Yorke, H. W.  \&  Bodenheimer, P., 1979, ApJ, 80, 110
\bibitem[Tikhonov \&   Samarskii (1990)]{Tikh90} Tikhonov, A.N. \&   Samarskii, A.A. 1990, Equations of Mathematical Physics (Dover Publications)
\bibitem[Walch et al. (2013)]{Wal13}  Walch S.,  Whitworth, A.P.,  Bisbas, T.G.,   Wunsch, R. \& Hubber, D.A., 2013, arXiv:1306.4317
\bibitem[Whalen \& Norman (2008)]{Whal08} Whalen D. \& Norman M. L., 2008, ApJ, 673, 664
\bibitem[Whitworth et al. (1994)]{Whith94} Whitworth, A. P., Bhattal, A. S., Chapman, S. J., Disney, M. J., \& Turner, J. A. 1994, A\&A, 290, 421
\bibitem[Zonenko \& Chernyi (2003)]{Zon03} Zonenko, S. I. \&  Chernyi, G. G., 2003, Doklady Physics, 48, 239
\end{thebibliography}
\end{document}